\documentclass[conference, a4paper]{IEEEtran}

\usepackage{url,hyperref}
\usepackage{mathtools}

\newcommand{\scum}{SC\textmu M}

\def\BibTeX{{\rm B\kern-.05em{\sc i\kern-.025em b}\kern-.08em
    T\kern-.1667em\lower.7ex\hbox{E}\kern-.125emX}}

\begin{document}

\title{A Python-based RTL Generator Demonstrated on a Low-IF 2-FSK Wireless Communication System}
\author{\IEEEauthorblockN{Brandon P. Hippe* and David C. Burnett**}
\IEEEauthorblockA{\emph{Department of Electrical and Computer Engineering, Villanova University, Villanova, PA 19085 USA}}
*bhippe@villanova.edu, **david.burnett@villanova.edu}
\maketitle

\begin{abstract}
Hardware optimization is critical in the design of efficient wireless communication systems. Wireless communication hardware often consumes a significant fraction of the total system's power budget, with much of this power used in circuits that reduce various types of noise, particularly in the analog front end. The Single-Chip Micro Mote \cite{scumpaper}, or \scum{}, uses a crystal-free radio architecture and makes design trade-offs that favor power consumption over noise performance while maintaining standards compatibility with popular Internet-of-Things (IoT) protocols such as IEEE 802.15.4 and Bluetooth Low Energy. In the continued development of \scum, we recognize that the digital baseband hardware developed can be more closely optimized with the architecture of the chip. In this paper, we present an extensible Python-based RTL generator that is closely linked to simulation and testing environments. This approach provides flexibility for use on different hardware platforms, such as tape-outs and FPGA implementations, and has promise in AI-assisted design workflows.
\end{abstract}

\begin{figure}[t]
    \centering
    \includegraphics[scale=.8]{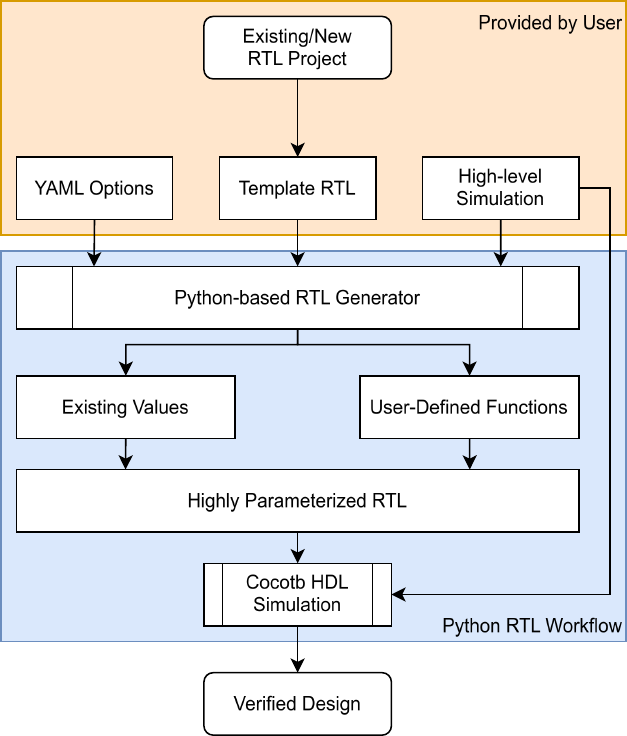}
    \caption{Flow diagram demonstrating the process of using the proposed RTL generator to parameterize a design.}\label{fig:flow}
\end{figure}

\section{Introduction}

Current crystal-free System-on-Chip (SoC) research aims to enable such devices to serve as standards-compatible transceivers using established narrowband FSK standards in the 2.4~GHz ISM band, such as Bluetooth Low Energy (BLE) and IEEE 802.15.4. BLE and IEEE 802.15.4 are ubiquitous IoT wireless protocols for low-power systems at relatively low data rates, whose specifications are written to be feasible to implement with low-cost hardware. \scum{} is an ideal platform for these protocols, as it is designed to enable fine-grained tuning to local oscillator, receiver gain and data recovery settings, and more. \scum{} is unique in its use of a crystal-free architecture, allowing \scum{} to approach a truly monolithic integration: The only external components it needs to operate are a power source and an antenna. \scum{} elides the use of power-hungry off-chip crystal oscillator references \cite{crystal-free}, instead utilizing entirely on-chip oscillators for its digital clock and RF signal generation. As a result, \scum{} can also forgo the use of a Phase-Locked Loop (PLL) in the RF Front End. However, these design choices come at the cost of reduced phase-noise performance and, therefore, degraded communication system performance. As a result, a low-IF baseband architecture is used instead of more conventional zero-IF architectures.  As shown in \cite{APCCAS}, careful analysis of simulations is crucial in reducing the impact of phase noise trade-offs made by the use of a crystal-free architecture, and we believe that a well-optimized design is capable of sufficient performance as an IEEE 802.15.4/BLE transceiver \cite{BLE_Oscs}.

Among the many challenges with crystal-free radio design is the difficulty of accurately simulating the RF front-end components to a reasonable degree \cite{rohail_best_2025}. Phase noise simulations can be extremely time and compute-intensive, and many of the common approaches have widely varying results. As such, the concept of a ``digital twin" is highly useful in crystal-free research: \scum{}'s existing RF front end can be used as a signal source and routed off-chip to a test digital component implemented in an FPGA. This makes pre-silicon validation much easier than in a simulation-only environment. During the design of 2-FSK receiver hardware, the need arose for the hardware to be utilized on varying platforms; namely, the FPGA on an ADALM-PLUTO Software Defined Radio \cite{ADALM-PLUTO}, an FPGA interfaced with \scum{}, and an upcoming tapeout. Each of these implementations requires a unique design to meet respective specifications. This generator framework was developed so that our existing 2-FSK baseband simulations could directly inform the hardware design and be used in testing and validation.

The rest of this paper details the functionality of the developed RTL generator for generalized use, along with describing its use cases in ongoing and future \scum{} development projects.

\section{RTL Generator}

The RTL generator framework described here is intended as a simple bridge from a Python-based simulation workflow to hardware environments using Verilog or SystemVerilog HDL, though the concepts used here apply to other languages. It is not intended to use Python to write HDL, but rather as a tool to make parameterizing HDL more extensible. Although this system was developed with the generation of 2-FSK Digital Baseband RTL in mind, it is theoretically usable for use in any RTL generation where a Python-based simulation is available.

\subsection{Framework Overview}

A flow diagram for this framework is shown in Figure \ref{fig:flow}. In principle, this framework takes a Verilog or SystemVerilog template and fills in various parameters using Python. It uses two types of input files: template RTL file(s) and an options YAML file.

Template RTL files are written the same way as any other RTL, with sections to be parameterized indicated using specific formats in comments or directly in the RTL. At runtime, any such strings are replaced with their corresponding parameter, or if none exists, the generator tries to call a function named \verb|parameter_name|. Such a function returns the RTL to be inserted at that position as a string. This functional approach allows for more extensible parameterization of the generated RTL. For example, in the example digital baseband generator, the multiplications necessary in the matched filter implementation are automatically converted into bit-shift and accumulate operations. This helps synthesis tools use DSP slices more efficiently and is also useful for tape-out synthesis. This approach is also what allows the digital baseband generator to work for an arbitrary choice of IF, as specific parameters in the clock recovery can be calculated using existing simulations. Furthermore, the flexibility of parameter specification enabled by this approach allows templates to be reusable and simple to make.

The options YAML file defines the default arguments for filling in the template or in a function that returns a string of RTL. This file is also used to create a simple command-line interface for the generator, allowing the user to override any argument at generation time.

Although this framework is not a high-level synthesis tool and still requires designers to write RTL, it has many benefits compared to a high-level synthesis approach.

\subsection{Use with existing RTL Projects}

Using this generator with an existing HDL project is simple. The project can be installed using \verb|pip|, and contains commands to set up or update a project and generate the RTL. The RTL generator can be run at any level of the hierarchy, and if the provided setup option is used, only a single YAML options file is needed, located at the top level of the hierarchy. This enables all relevant sub-modules to be generated using the same parameters, simplifying the process of adding additional blocks to the generator flow, and has the added benefit of putting all default options in a single location. After generation, the generator prints a table of the values used in the generation of that specific module as a comment at the top of the file, allowing for these values to be read and used in later steps. This table is also properly formatted in Markdown such that auto-documentation tools such as \cite{noauthor_teroshdl_nodate} can include this in generated documentation.

\subsection{Integration with Simulation}

One of the most interesting use cases of a generator flow such as this is in hardware optimization. Direct integration with Python simulations means that some parameters could be optimized and determined in the simulation rather than specified by the user. One can easily imagine the use of various machine learning approaches to do this optimization, which would speed up the design process in comparison with the extensive parameter space search performed for this digital baseband design. For example, when examining IF choices for matched filter performances, it was found that certain intermediate frequencies lead to exceedingly simple templates without sacrificing the performance of the matched filter. There are likely many more simplifications available in this design, which would directly lead to lower hardware utilization and, therefore, power consumption and cost savings.

\subsection{Integration with Verification}

Another key benefit of this framework is its usefulness in hardware verification. Cocotb \cite{cocotb} is a Python-based verification tool that is perfect for use with this generator framework. The parameter tables were added specifically so that the RTL parameters could be read from Cocotb verification scripts and allow the same Python simulations to be used in verification as well. This is crucial to verify that the RTL functions in the same way as simulations, allowing designers to be confident that the design will perform as expected when synthesized into real hardware.

\section{Conclusion}

In this paper, we propose a Python-based RTL generator framework that bridges the gap between high-level simulations and hardware designs. The described approach allows RTL to be easily modified for different hardware platforms and has promise in being used for AI-assisted design optimization. The framework is demonstrated in the generation of digital baseband hardware optimized for crystal-free wireless communication systems using 2-FSK protocols, such as Bluetooth Low-Energy and IEEE 802.15.4. This generator has helped produce hardware used on a ``digital-twin" prototyping FPGA platform and for hardware that will be included in an upcoming tapeout.

\section*{Appendix I}
The RTL generator code described is available at \url{https://github.com/burnettlab/rtl-generator} or \url{https://pypi.org/project/rtl-generator}.

\bibliographystyle{IEEEtran}
\bibliography{references}

\end{document}